\documentclass[prd, twocolumn, nofootinbib, floatfix]{revtex4-1}

\usepackage{amsmath}
\usepackage{graphicx}
\usepackage{dcolumn}
\usepackage{bm}
\usepackage{epsfig}
\usepackage{amssymb,latexsym,mathrsfs}
\usepackage{graphicx}
\usepackage{color}
\usepackage{hyperref}

\hypersetup{
    colorlinks=true,
    linkcolor=red,
    citecolor=blue,
} 

\newcommand{\be}{\begin{equation}}
\newcommand{\ee}{\end{equation}}
\newcommand{\bs}{\begin{split}} 
\newcommand{\bea}{\begin{eqnarray}}
\newcommand{\eea}{\end{eqnarray}}

\newcommand{\om}{\Omega_m}

\newcommand{\gm}{G_{\rm matter}} 
\newcommand{\dgm}{\delta G_{\rm matter}}

\newcommand{\dg}{\delta G}

\newcommand{\fs}{f\sigma_8}
\newcommand{\sig}{\sigma}
\newcommand{\dchi}{\Delta\chi^2}

\begin{document}

\title{Subpercent Accurate Fitting of Modified Gravity Growth} 

\author{Mikhail Denissenya${}^1$ and Eric V.\ Linder${}^{1,2}$} 
\affiliation{${}^1$Energetic Cosmos Laboratory, Nazarbayev University, 
Astana, Kazakhstan 010000\\ 
${}^2$Berkeley Center for Cosmological Physics \& Berkeley Lab, 
University of California, Berkeley, CA 94720, USA} 

\begin{abstract} 
Adding to our previous method for dealing with 
gravitational modifications at redshift $z\gtrsim3$ through a single 
parameter, we investigate treatment of lower redshift modifications to 
linear growth observables. We establish subpercent accurate fits to the 
redshift space distortion observable $f\sigma_8(a)$ using 
two parameters binned in redshift, testing the results for modifications 
with time dependence that rises, falls, is nonmonotonic, is multipeaked, 
and corresponds to $f(R)$ and braneworld gravity. The residuals are then 
propagated to cosmological parameter biases for DESI observations, and 
found to cause 
a shift in the dark energy joint confidence contour by less than the 
equivalent of $\sim0.1\sigma$. The proposed 2--3 parameter modified gravity 
description also can reveal physical characteristics of the underlying theory. 
\end{abstract}

\date{\today} 

\maketitle

\section{Introduction} 

Gravity has been stringently tested on terrestrial, solar system, and 
astrophysical scales but cosmic scales represent a further $10^6$ to 
$10^{14}$ extrapolation in length scale (from galaxy scales or solar 
system scales respectively). Given that cosmic expansion is, surprisingly, 
accelerating, opposite to the expectation from gravity acting on matter, 
it is natural to desire tests of gravity on cosmic scales. 

The cosmic expansion by itself cannot distinguish between a change in 
the laws of gravity and in the material contents, i.e.\ a dark energy 
component, but the combination with the cosmic growth of large scale 
structure can. Therefore considerable effort has gone into understanding 
the effect of modified gravity on cosmic structure growth; for reviews, 
especially model independent work, see 
\cite{review1,review2,1604.01059,1703.01271}. 

Numerous alternatives to general relativity exist, with many of them 
falling within the Horndeski class of gravity, or described by an effective 
field theory approach (see \cite{review3} and references therein). These 
approaches involve four or more free functions of time, in addition to 
the expansion history, with no prescription for how they should behave. 
Even next generation data will not be able to constrain four functions, 
or more than a few parameters. Simple functional forms tend to be highly 
restrictive and possibly poor approximations \cite{1512.06180,1607.03113}. 
Thus we must either work one at a time with one particular 
model of gravity, one particular functional form within that model, and 
one particular parameter set within that functional form (e.g.\ $f(R)$ 
gravity, of the Hu-Sawicki \cite{husawicki} form, with $n=1$), or seek a 
phenomenological low dimensional model independent approach. 

If we follow the data, then in the subhorizon, quasistatic limit (applicable 
to where precision data will lie) the linear growth of structure is determined 
by a generalized Poisson equation, as clearly shown by \cite{bz}. Here the 
gravitational strength determining matter density perturbation growth is 
$\gm(k,a)$ rather than Newton's constant $G_N$, where the scale factor $a$ 
represents 
the time dependence and the wavenumber $k$ the scale dependence. This is a 
robust treatment for modified gravity under these circumstances \cite{bz}. 
Thus the issue, if one is concerned with using cosmic growth data to 
test gravity, is how to parametrize $\gm$. 

One advance in this direction appears in \cite{paper1} (hereafter Paper 1). 
The authors derived, and demonstrated numerically, that modifications to the 
gravitational strength at early times, $a\lesssim0.25$ in the matter dominated 
era, could be modeled with high accuracy in their effects on the growth 
observables by a single parameter, $G_{\rm hi}$, related to the effective 
area under the $\dgm(a)$ curve. This is accurate to 0.3\% or better. The 
treatment of later time modifications to gravity, however, was left as an 
unresolved question. The aim here is to address it. 

In Sec.~\ref{sec:model} we describe the variety of gravity models that we 
seek to fit, and the model independent method used. The specific approach 
and observational data used is described in Sec.~\ref{sec:method}, and 
the results for the accuracy of the parametrization are presented in 
Sec.~\ref{sec:results}. We propagate the fitting residuals to cosmological 
parameter bias in Sec.~\ref{sec:bias}. Section~\ref{sec:discuss} discusses 
how to use the parametrization with data in a practical sense, and how to 
extract key properties of the gravity theory from the results. We conclude 
in Sec.~\ref{sec:concl}.

\section{Gravity Models and Fits} \label{sec:model} 

In the quest for a low dimensional parametrization of the effect of 
modified gravity on linear growth observables, we want not only an 
accurate parametrization but a broadly model independent one. Functional 
forms such as power laws tend to be limited, and often bias the results 
by weighting unfairly parts of the cosmic history,  
as well as the results being sensitive to the power law, or prior on the 
power law, assumed, while being unable to constrain the power law well 
\cite{1109.1846,1612.00812}. Assuming a close relation with the effective 
dark energy density also yields misleading conclusions 
\cite{1512.06180,1607.03113}, with the simplest counterargument being that 
$f(R)$ gravity often shows a gravitational strength that only deviates from 
general relativity at quite late times, e.g.\ reaching 1\% only at 
$z\approx1.5$, when the dark energy density fraction is already greater than 15\%. 
Conversely, assuming that gravitational modifications only occur at late 
times can miss important aspects of many theories such as the Horndeski 
class of gravity. 

Therefore we turn toward bins in scale factor or redshift as a model 
independent approach. These have 
been successfully used in projecting future constraints on modified 
gravity, e.g.\ \cite{1212.0009,1612.00812}. We will lay out a methodology 
for deciding on the number of bins, and interpreting the meaning of the 
results in the remainder of the article. We emphasize that our goal is 
to fit to the observables, specifically the redshift space distortion (RSD) 
function $\fs(a)$, not the theory function $\gm(a)$. 

To test the efficacy of binned parameters, we have to compare them 
to some underlying ``true'' theory. 
To robustly explore the comparison of the results of the binned model with 
the exact theory, we need to ``stress test'' the binned approximation by 
comparing it to a wide variety of theoretical behavior. Since our focus 
is on growth observables and looking for signatures of modified gravity, 
we use identical expansion histories for the model and the theory case 
it is attempting to fit. 

The theory behaviors should be fairly 
realistic, with enough complexity and features to provide an adequate 
test of the binned parametrization. We adopt six different forms of 
scale factor dependence to test: 
\begin{enumerate} 
\item a nonmonotonic function, taken to be a Gaussian of variable width and 
location, as in Paper 1, but at late times; 
\item a rising function; 
\item a falling function (it is obvious that the 
constant function considered in Paper 1 can be fit by a binned 
parametrization); 
\item a multipeaked function such as seen in some Galileon gravity cases 
(e.g.\ see Fig.~3 of \cite{1607.03113}), taken to be a sum of Gaussians; 
\item braneworld theory given by DGP gravity \cite{dgp1,dgp2}; 
\item $f(R)$ gravity. 
\end{enumerate} 

The Gaussian deviation, normalizing $\gm$ by Newton's constant so 
that general relativity has $\gm=1$, is 
\be 
\dgm=\dg\,e^{-(\ln a-\ln a_t)^2/(2\sig^2)} \ , 
\ee 
where we will study the results for various central values $a_t$ and widths 
$\sig$. 

The rising parametrization is 
\be 
\dgm=\dg\,a^s\quad {\rm for}\ a>a_\star\ , 
\ee 
and otherwise zero, where $a_\star$ is a cutoff scale factor. 
We might choose $a_\star=0.25$ ($z=3$) since from Paper 1 we know how 
to treat the deviations for $z>3$. 
The falling parametrization is 
\be 
\dgm=\dg\,a^{-s}\quad {\rm for}\ a>a_\star\ , 
\ee 
and otherwise zero. 

The sum of two Gaussians gives either a multipeaked function or a broader 
deviation, depending on the separation of the Gaussians and their width. 
We take $a_t=0.3$ and $a_t=0.7$, with either $\sig=0.25$ (giving multiple 
peaks) or $\sig=0.5$ (giving a broad deviation). 

For DGP gravity, the expansion history is given by the modified 
Friedmann equation 
\be 
\frac{H(a)}{H_0}=\frac{1-\om}{2}+\sqrt{\frac{(1-\om)^2}{4}+\om\,a^{-3}} \ , 
\ee 
and the modified gravity strength is 
\be 
\dgm=-\frac{1}{3}\,\frac{1-\om^2(a)}{1+\om^2(a)} \ . 
\ee 
where $\om(a)=\om\,a^{-3}/[H(a)/H_0]^2$. At early times, $\om(a)\to1$ and 
the strength restores to the Newtonian value, i.e.\ $\dgm=0$. In 
the asymptotic future, the Hubble parameter freezes to a de Sitter state, 
$H/H_0\to 1-\om$ and gravity freezes to $\dgm=-1/3$, i.e.\ $\gm=2/3$, 
weaker than Newtonian due to the extra dimensional leakage. 

For the $f(R)$ scalar-tensor gravity case, we adopt exponential gravity. 
See \cite{0905.2962} for the relevant equations; the basic features are 
that the expansion history is close to $\Lambda$CDM but with the dark 
energy equation of state varying slowly around $w=-1$, on both the phantom 
and normal sides. The gravitational strength is greater than Newtonian, 
rising from the general relativity value at high redshift (and indeed 
for $z\gtrsim1.5$) to 4/3 times the value; i.e.\ $\dgm$ goes from 0 
to 1/3. 

These are all compared to the results from the binned parametrization. 
This is simply $\gm$ piecewise constant in 
two bins of $a$. These span $a=[0.25,0.5]$ and $a=[0.5,1]$, since these 
are the main observational windows. As discussed in the next section, 
if the data show the need then we include an early time parameter 
corresponding to the area parameter of Paper 1, implemented as a constant 
value in a window $a=[0.1,0.25]$. We smooth the bin edges with a tanh 
function; results are insensitive to a smoothing width below 
$\Delta\ln a=0.01$.

\section{Method and Data} \label{sec:method} 

For the theory and the binned model we solve the growth evolution 
equation using a fourth order Runge-Kutta method. The background expansion 
is taken to be a flat, $\Lambda$CDM cosmology with $\om=0.3$, except for 
the DGP and $f(R)$ gravity cases where we simultaneously solve the background 
evolution equations. 
We then compare the observable RSD quantity $\fs(a)$ between 
the input theory and the binned parametrization results and determine the 
maximum and rms deviation. 

The bin values are optimized by minimizing 
one of these quantities. We find that substantially similar values result 
from either optimization. For values used below, we nominally minimize the 
maximum deviation over the range $z=0.15$--1.9, corresponding to the 
data used, as discussed below. 

Note that a point deviation value, or rms, is not really the key quantity. 
Neither will pick up particular trends, such as the RSD observable being 
high for several redshift bins, then low for several, as opposed to random 
scatter. One possibility is to use some statistic such as the crossing 
statistic \cite{crossing1,crossing2} that does identify such patterns. 
However, what 
we are really interested in is the propagation of the residual between the 
theory prediction and the binning approximation to the cosmological 
parameters. For example, even a moderately large amplitude high frequency 
oscillation will not affect the cosmological determination since it does 
not look like a shift in cosmology. Therefore we use the maximum deviation 
in Sec.~\ref{sec:results} to determine the bin values, but then propagate 
the residuals to cosmology in Sec.~\ref{sec:bias} with the Fisher bias 
formalism. 

As our mock data we take RSD measurements as projected for the Dark Energy 
Spectroscopic Instrument (DESI \cite{desi}), with the uncertainties on 
$\fs(a)$ given in Tables~2.3 and 2.5 of \cite{desitable}, for 
$k_{\rm max}=0.1\,h/$Mpc.

\section{Results} \label{sec:results} 

\begin{figure}[htbp!]
\includegraphics[width=\columnwidth]{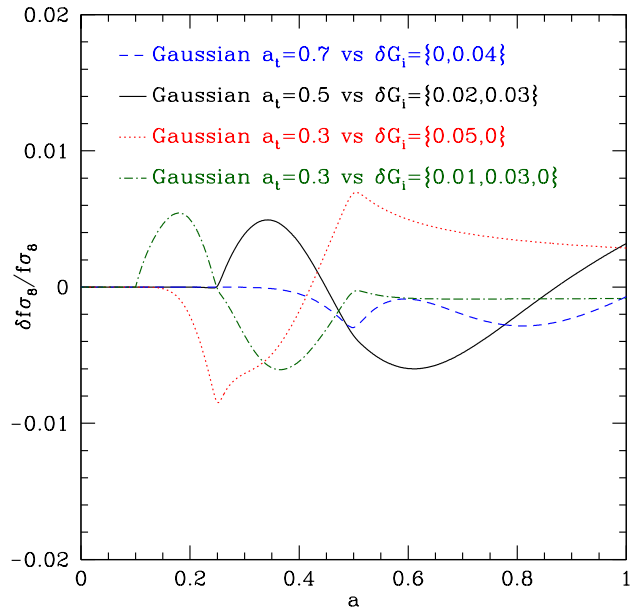} 
\caption{
The accuracy of fitting the observational RSD factor $\fs$ with two late 
time bins for modified gravity $\dg(a)$ is compared to that for the exact 
theory case. The theory model has a Gaussian $\dg(a)$ with parameters 
$\dg=0.05$, $\sig=0.25$, and $a_t=0.3$ (dotted red), 0.5 (solid black), 
0.7 (dashed blue). The dot dashed green curve shows the $a_t=0.3$ case 
fit when allowing for a third, early bin due to the early modification. 
} 
\label{fig:gaus1} 
\end{figure}

For theory Case 1, with nonmonotonic time dependence, we adopt Gaussian 
modifications $\dgm$ with amplitude $0.05$ and width $\sig=0.25$. 
Figure~\ref{fig:gaus1} shows the deviations in $\fs(a)$ for the theory models  
with $a_t=0.3$, 0.5, 0.7 vs the binned approximations.

We see that two bin parameters achieve subpercent level residuals relative 
to the exact results over the full redshift range. For the $a_t=0.3$ case, 
the modification extends earlier than the bin start at $a=0.25$. If we 
wanted to add the early time modification area parameter, or equivalently 
a third bin at $a=[0.1,0.25]$, we reduce the maximum deviation from 0.9\% 
to 0.6\% (though nearly 0 for $a>0.5$). There is no particular need for 
a third parameter even in this case, and this conclusion is verified by 
the cosmology bias analysis in Sec.~\ref{sec:bias}. 

\begin{figure}[htbp!]
\includegraphics[width=\columnwidth]{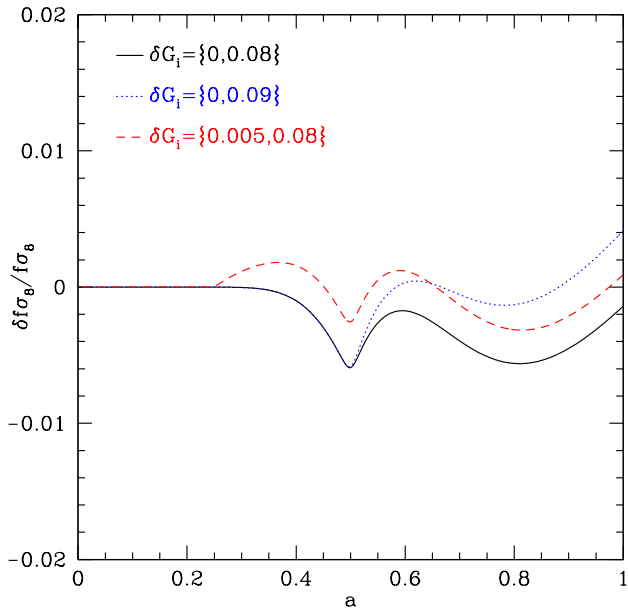}
\caption{
The Gaussian modification case with larger amplitude $\dg=0.1$, 
$a_t=0.7$, $\sig=0.25$ can still be accurately fit with two bins. 
Three different possibilities are shown, with different rms residuals, 
but all give high accuracy fits. 
}
\label{fig:hiamp}
\end{figure}

Note that for modifications close to the present, e.g.\ the $a_t=0.7$ case, 
even just one parameter, from the bin $a=[0.5,1]$ gives excellent results. Even 
if we double the amplitude, to $\dg=0.1$, the maximum deviation in $\fs$ 
stays under 0.5\% as seen in Fig.~\ref{fig:hiamp}. This also illustrates 
the possibility of trading off a residual curve that stays closer to 0 
for much of its run, but has an overall larger max--min range, with one that 
is further from 0 but rather flat. We might expect that the latter, though with 
greater rms deviation, has less cosmological consequence, and indeed this 
holds true. One could further improve on the fit by allowing the second 
bin to enter, and then the high amplitude case has only 0.2\% maximum 
deviation.

\begin{figure}[htbp!]
\includegraphics[width=\columnwidth]{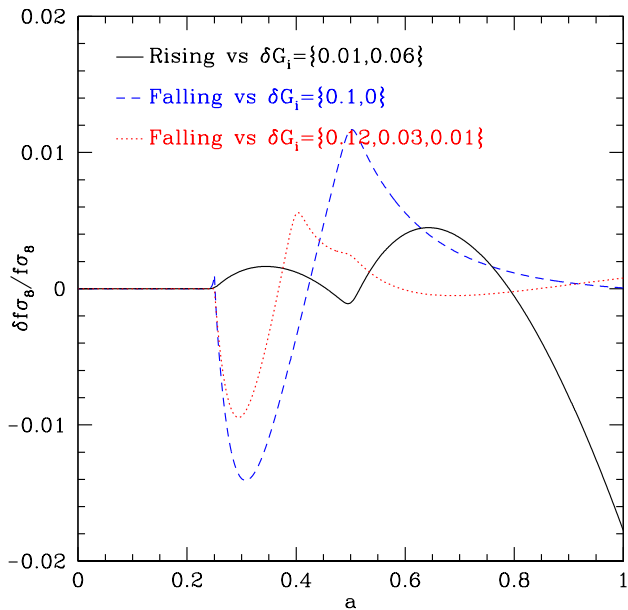}
\caption{
Modifications that rise or fall monotonically over the range $a=[0.25,1]$ 
can also be fit well with just two parameters, though the less realistic 
falling case benefits from splitting the $a=[0.25,0.5]$ bin. 
}
\label{fig:fallrise}
\end{figure}

Turning to theory Case 2 and 3, we consider power law rising and falling 
time dependences, with $\dg\propto a^3$ and $a^{-3}$. 
(Note that the parametrizations used by 
\cite{1703.10538,1705.04714} are basically within the rising class.) 
The 
normalization we use gives a maximum $\dg=0.21$, a considerable amplitude, 
with the rising case reaching this at $a=1$ and the falling case at its 
starting point $a=0.25$. As seen in Fig.~\ref{fig:fallrise}, the rising case 
can be easily fit with two bin parameters, and the maximum deviation is 
less than 0.5\% for $a<0.85$. 
This is more than satisfactory as the 
DESI data projects an uncertainty of greater than 12\% for $a>0.85$ due 
to the small cosmic volume available. (Better measurements may be possible 
by using peculiar velocities \cite{howlett}.) In any case, one could 
achieve 0.9\% accuracy over all $a$ using $\dg_3=0.075$ instead of 0.06.) 

The falling case achieves 1.4\% accuracy with two bins, due to its large 
amplitude and rapid variation in the bin $a=[0.25,0.5]$. This deviation 
pattern would 
be noticeable in the data fits, and would spur an analysis where this bin 
would be split in two, e.g.\ $a=[0.25,0.4]$ and $a=[0.4,0.5]$. With this 
three parameter fit, the residuals obtain subpercent accuracy. Either way, 
this sort of oscillation in residuals does not tend to give a cosmology 
bias, and thus is mostly harmless. Finally, note that 
in any case such a falling model is not generally seen in gravity theories 
commonly investigated.

\begin{figure}[htbp!]
\includegraphics[width=\columnwidth]{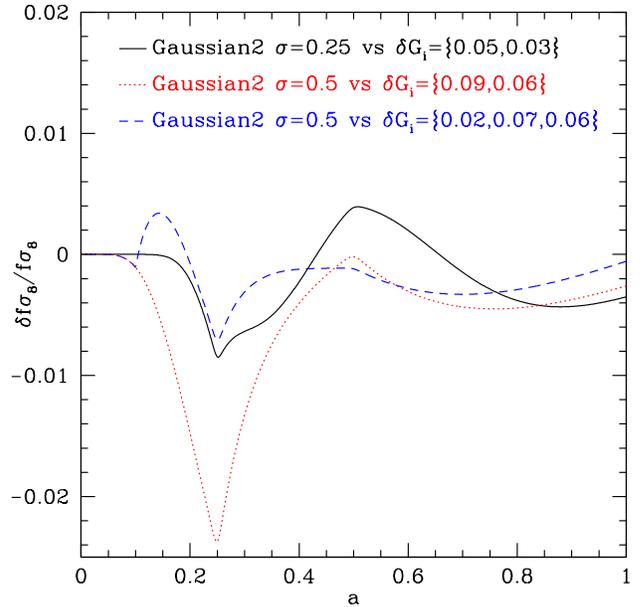}
\caption{
The multipeak (two Gaussians) model can have subpercent residuals when using two 
bin parameters. When the early Gaussian has substantial support at $a<0.25$ 
(the $\sig=0.5$ case) then adding a third, early bin substantially improves 
the accuracy. 
}
\label{fig:gaus2}
\end{figure}

For the multipeak case, reminiscent of modifications seen in theories 
with many terms such as Horndeski gravity, we model this by the sum 
of two Gaussians, at $a_t=0.3$ and 0.7. We adopt $\sig=0.25$ to obtain 
a multipeak $\dg(a)$, and also investigate $\sig=0.5$ 
to give a broad, non-Gaussian $\dg(a)$. Figure~\ref{fig:gaus2} shows the 
accuracy of the binned approximation.

For broad early modification, one needs an early bin for subpercent accuracy, 
i.e.\ the area parameter of Paper 1. As discussed in Sec.~\ref{sec:discuss}, 
the need for an early bin makes itself known from the trend of data points 
with redshift. However, if the precision data extends only to $z\approx1.9$ 
($a\approx0.34$) then two bins gives 0.8\% precision. 

\begin{figure}[htbp!]
\includegraphics[width=\columnwidth]{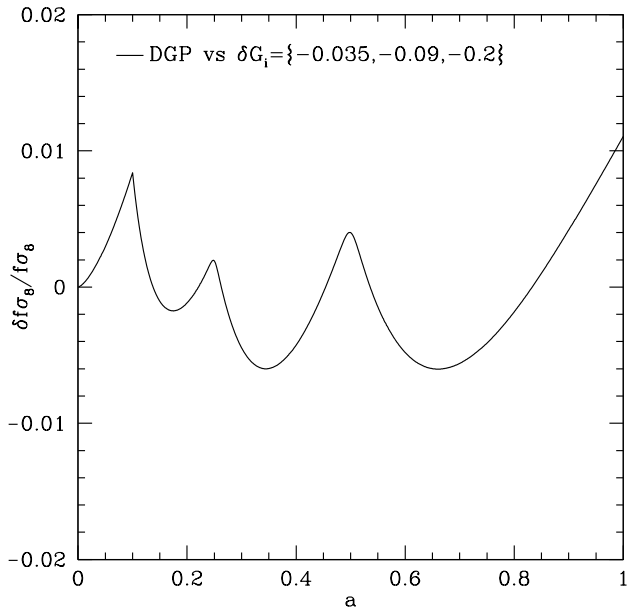}
\caption{
DGP gravity is well fit by two bin parameters plus the early time 
modification, or area, parameter. 
}
\label{fig:bwsimul}
\end{figure}

Finally, we consider actual gravity theories. 
Braneworld gravity, specifically DGP gravity, exhibits a significant change 
in the strength of gravity, with $\dg\approx -1/3$. Since its deviation from 
general relativity starts relatively early, i.e.\ once $\om(a)$ starts to 
deviate from 1, we expect to need to include the third, area parameter or 
early bin. The results appear in Fig.~\ref{fig:bwsimul}.

Fitting to binned $\dg$ gives an oscillating residual, reflecting that 
$\dg_{\rm DGP}$ is quite smooth and monotonic so each bin fits the average 
value within its redshift range, under- and overestimating the function in 
the different halves of the bin. The amplitude of the residuals is 
0.6\% except at very early or late times. (Again, the DESI measurement 
precision at $a<0.2$ or $a>0.9$ is such that even a 1\% residual there 
needs no improvement.) Since this oscillatory pattern 
does not look like cosmological parameter variation, we expect little bias in 
the three bin case. 

\begin{figure}[htbp!]
\includegraphics[width=\columnwidth]{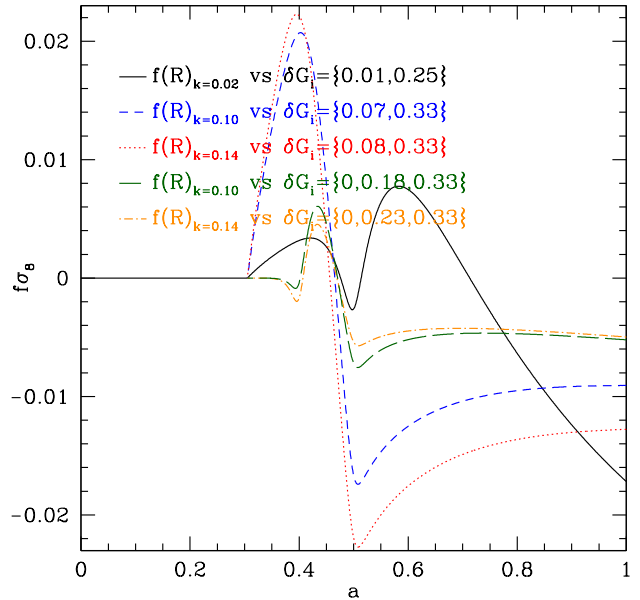}
\caption{
Exponential $f(R)$ gravity is fit by 2-3 bin values, with different 
gravitational strengths at different wavenumbers $k$ (i.e.\ scale 
dependent). Due to the steepness of the time evolution of the modification, 
the fit is greatly improved when using a third bin made by splitting the 
$a=[0.25,0.5]$ bin (long dashed green and dot dashed orange curves, relative 
to dashed blue and dotted red curves). 
}
\label{fig:frsimul}
\end{figure}

Finally, consider $f(R)$ theory. We adopt the exponential gravity form, 
with $c=3$, which is consistent with observations \cite{0905.2962}. Recall 
that $f(R)$ gravity also exhibits a significant change
in the strength of gravity, with $\dg\approx 1/3$. It generally has a steep 
time dependence, with $\dg(a)=0$ until quite recent times and then rapidly 
rising. For example, it reaches 1\% deviation from general relativity at 
$z\approx 1.5$ and has 33\% deviation at $z=0$. In addition, the gravitational 
strength, and hence growth, is scale dependent. Figure~\ref{fig:frsimul} 
shows the binned gravity values for growth at three separate wavenumbers $k$.

If we knew the true theory was $f(R)$ gravity then we could scale the 
bin values according to the predicted scale dependence of $\gm$ in $f(R)$ 
theory, i.e.\ $[3+4k^2/(aM)^2]/[3+3k^2/(aM)^2]$ where $M(a)$ is the 
scalaron mass \cite{bz,0511218,0709.0296}. 
However, we do not know this a priori. (See \cite{1707.08964} for a more 
model independent approach.) Indeed, we discover the scale 
dependence empirically, when we compare the data to the result from the 
binned gravity fit and the residuals indicate a discrepancy that can be removed 
by introducing different binned values for different wavenumbers. Note, 
however, that the binned values are fairly similar for $k\gtrsim 0.1\,h$/Mpc. 

The steepness of the evolution of the gravitational strength shows up 
not only in the two-bin values, but in the strong improvement made when 
splitting the $a=[0.25,0.5]$ bin into two parts, $a=[0.25,0.4]$ and 
$[0.4,0.5]$. As mentioned above, there is almost no deviation from 
general relativity for $a<0.4$, and the finer early bin value is consistent 
with zero, while the larger, later split bin value greatly reduces the 
residuals. 

As discussed in Sec.~\ref{sec:discuss}, the steepness of the increase in 
the bin values for any $k$, the late time value near $1/3$, and the scale 
dependence would together allow us to deduce -- from a model independent 
analysis method! -- that the true gravity theory is likely of the $f(R)$ 
class.

\section{Impact on Cosmology} \label{sec:bias} 

We established in the previous section that the residuals from fitting 
the RSD observable $\fs(a)$ with the two or three gravity parameters are 
at subpercent accuracy. Since next generation, DESI precision for $\fs(a)$ 
will be at the $\gtrsim 2\%$ ($\gtrsim 1\%$ if we used data out to 
$k_{\rm max}=0.2\,h$/Mpc), this seems sufficient. However, if the residuals 
coherently combine in their effect, due to a time dependence mimicking a 
shift in a cosmological parameter, they have the potential to bias the 
cosmological conclusions. 

Therefore we now propagate the residuals in $\fs(a)$ to the cosmological 
model parameters. We use the Fisher bias formalism to carry this out 
\cite{fisbias}. The set of cosmological parameters considered is 
the present matter density in units of 
the critical density, $\om$, the dark energy equation of state parameter 
today $w_0$, and a measure of its time variation $w_a$, 
where $w(a)=w_0+w_a(1-a)$. We use the DESI $\fs(a)$ data as described in 
Sec.~\ref{sec:method} and to break background degeneracies 
we apply a simple Gaussian prior on the matter density $\sigma(\om)=0.01$. 
Our fiducial model is flat $\Lambda$CDM with $\om=0.3$. 

The bias on a parameter $p_i$ due to a misestimation of the observable 
$\Delta O(a)$ is (see, e.g., \cite{0604280}, including for the case where 
the error matrix is not diagonal) 
\be 
\delta p_i=\left(F^{-1}\right)_{ij}\sum_k \frac{\partial O_k}{\partial p_j} 
\frac{1}{\sigma_k^2}\,\Delta O_k \ , 
\ee 
where $O_k$ is the $k$th observable (i.e.\ $\fs(z_k)$) and $F$ is the 
Fisher matrix. 

Once we have the set of $\{\delta p_i\}$ we can quantify the bias statistically. 
One way is of course simply looking at $\delta p_i/\sigma(p_i)$, the bias 
relative to the statistical uncertainty. A common statistical quantity that 
employs this is the risk, which take the square root of the quadrature sum 
of the bias and dispersion. We can take the ratio of the risk to the 
statistical uncertainty to find the bloat, or effective increase in 
the uncertainty on a parameter: 
\be 
B_i=\frac{\sqrt{\delta p_i^2+\sigma^2(p_i)}}{\sigma(p_i)}=
\sqrt{1+[\delta p_i/\sigma(p_i)]^2} 
\ee 
This quantity appears for example in the Rao-Cram{\'e}r-Frechet bound \cite{rao}. 

Finally, perhaps most useful is the shift induced in the joint parameter 
fitting, e.g.\ in the offset of the derived values from the true best fit 
in the dark energy equation of state plane $w_0$--$w_a$. The shift relative 
to the likelihood contours at some confidence level presents an informative, 
quantitative assessment of the bias that takes into account parameter 
degeneracies. This is given by \cite{0508296,0812.0769} 
\be 
\Delta \chi^2=\sum_{ij} \delta p_i\,F^{({\rm red})}_{ij}\,\delta p_j \ , 
\ee 
where the reduced Fisher matrix $F^{({\rm red})}$ runs over only those 
parameters $p_i$, $p_j$ whose bias we are interested in, e.g.\ $w_0$ and 
$w_a$ for the 2D joint likelihood contour plot in the $w_0$--$w_a$ plane, 
and is marginalized over all others. 
This quantity automatically takes into account the {\it direction\/} of the 
shift, i.e.\ that a bias perpendicular to the degeneracy direction is more 
damaging than one along the degeneracy direction. 

Table~\ref{tab:chi} presents the values of $\Delta\chi^2$ for the joint 
$w_0$--$w_a$ likelihood, the maximum $\delta p/\sigma(p)$ for any of the 
cosmological parameters, and the maximum bloat in any of the cosmological 
parameters. Note that a shift of $\Delta\chi^2=2.3$ moves the true values 
out to the 68\% confidence contour, i.e.\ a joint $1\sigma$ bias. A shift 
smaller than this lies within the contour.

\begin{table}[htbp]
\begin{center}
\begin{tabular*}{\columnwidth} 
{@{\extracolsep{\fill}} l c c c }
\hline
Model & $\Delta\chi^2$ & [$\delta p/\sigma(p)$]$_{max}$ & Risk$_{\rm max}$ \\ 
\hline
Gaussian ($a_t=0.7$) & 0.02 & 0.02 & 1.00 \\ 
Gaussian ($a_t=0.5$) & 0.13 & 0.33 & 1.05 \\ 
Gaussian ($a_t=0.3$) & 0.16 & 0.22 & 1.02 \\ 
Gaussian ($a_t=0.7$; $\delta G=0.1$) & 0.09 & 0.04 & 1.00 \\
Gaussian$_3$ ($a_t=0.3$) & 0.09 & 0.22 & 1.02 \\  
Gaussian$^2$ ($\sigma=0.25$) & 0.03 & 0.09 & 1.00 \\ 
Gaussian$^2$ ($\sigma=0.5$) & 0.04 & 0.04 & 1.00 \\ 
Gaussian$_3^2$ ($\sigma=0.5$) & 0.03 & 0.07 & 1.00 \\  
Rising $a^3$ & 0.01 & 0.09 & 1.00 \\ 
Falling $a^{-3}$ & 0.36 & 0.25 & 1.03 \\ 
Falling$_{3s}$ $a^{-3}$ & 0.10 & 0.23 & 1.03 \\ 
DGP & 2.28 & 0.45 & 1.10 \\ 
DGP$_3$ & 0.00 & 0.02 & 1.00 \\ 
$f(R)$ ($k_0=0.02$) & 0.07 & 0.06 & 1.00 \\ 
$f(R)$ ($k_0=0.10$) & 1.81 & 1.34 & 1.67 \\ 
$f(R)_{3s}$ ($k_0=0.10$) & 0.18 & 0.40 & 1.08 \\ 
$f(R)$ ($k_0=0.14$) & 2.57 & 1.52 & 1.82 \\ 
$f(R)_{3s}$ ($k_0=0.14$) & 0.12 & 0.31 & 1.05 \\ 
\end{tabular*}
\caption{Parameter bias levels corresponding to the binned approximation 
of $\dg(a)$. $\Delta\chi^2$ is the shift in the dark energy equation of state 
parameter $w_0$-$w_a$ plane due to the bias; recall that $\Delta\chi^2=2.3$ 
corresponds to a $1\sigma$ shift in the joint parameter fit. The maximum 
bias of a parameter relative to its statistical uncertainty is shown 
in the $\delta p/\sigma(p)$ column. The Risk column shows the maximum 
``bloat'' of the Risk, i.e.\ the increase in the uncertainty due to the bias. 
The subscript 3 denotes the three bin fit with an early bin, and $3s$ 
denotes a three bin fit splitting the mid $z$ bin. The superscript 2 
denotes a convolution of two Gaussians, with $a_t=0.3$ and $a_t=0.7$. 
Note the approximate form can be good to $\Delta\chi^2<0.18$ for all models. 
} 
\label{tab:chi}
\end{center}
\end{table}

We see that in the results for the whole range of gravity models, using 
two parameters to represent the bin values, or in rare cases three 
parameters, keeps $\dchi<0.18$, i.e.\ less than a tenth of the distance 
to the $1\sigma$ joint likelihood contour. Alternately, the risk bloat factor 
is less than 1.08, i.e.\ the binned approximation only blows up the error 
bars, taking into account the systematic offset, by at most 8\%. Thus the two, or 
if needed three, parameter description of gravitational strength 
modification is statistically extremely robust.

\section{Observational Signatures} \label{sec:discuss} 

For any parametrization it is important that it be clear how it can 
be used to understand the data. That is, it should be of practical use 
to the observers and data analysts, as well as offering guidance to 
theorists.  

The binned parametrization is simple to apply, readily able to calculate 
$\fs(a)$ or other growth quantities with excellent accuracy. The steps in 
using it are straightforward: 
\begin{enumerate} 
\item Fit the predictions from two bins in 
$a=[0.25,0.5]$ and $a=[0.5,1]$ to the data. If all values are consistent 
with zero then general relativity is a viable gravity theory. If some 
values differ from zero with statistical significance, this is an alert 
that a potential signature of modified gravity has been found. 
\item If there are any residuals 
that show a pattern of exceeding the data error bars in some redshift 
range, then 
\begin{enumerate} 
\item Add the area parameter or equivalently a third bin at 
$a=[0.1,0.25]$ if the deviation shows up from early times (note the 
kink deviation and then slope in the residuals shown in the figures for 
the early Gaussian and DGP figures), or 
\item Split 
the $a=[0.25,0.5]$ bin into two bins over $a=[0.25,0.4]$ and $a=[0.4,0.5]$ 
if the deviation peaks in that range (see the falling and $f(R)$ figures). 
\end{enumerate} 
\item If the residuals indicate an overall poor fit, and in particular 
if the time evolution also looks steep (as in the $f(R)$ case), try 
separating the data into low 
and high wavenumbers to look for scale dependence. 
\end{enumerate} 
One could carry the bin refinement 
to a further level but none of the varied models we have considered require 
more than three bins, with two bins always sufficient here if data were at 
the 2\% precision level. 

The next question is how to interpret the results in terms of gravity 
theory. Note that the bin values are not a map of $\dgm(a)$ per se; they 
are a combination of the gravitation strength, the redshift weighting 
of the data and its precision, and a delay due to the convolution windowing 
of $\dgm(a)$ in the integral for $\fs(a)$. That said, they do provide a 
coarse guide to $\dgm$. 

A late bin value near $1/3$ inspires closer examination in terms of 
scalar-tensor gravity, while $-1/3$ would recall DGP gravity. The 
need for an early bin might lead to theories with early modifications 
such as the many members of the Horndeski class. Steepness of evolution 
in the binned value, reflecting steep time evolution of $\gm$, could point 
to $f(R)$ gravity, especially if splitting the $a=[0.25,0.5]$ bin led 
to a significant improvement in the residuals. And of course scale dependence 
gives theoretically important information. Thus, even though the analysis 
method is model independent, not assuming any functional form or even 
that gravitational modification is only a late time phenomenon, we can 
obtain substantial information about the theory characteristics from the 
signatures in cosmic growth data.

\section{Conclusions} \label{sec:concl} 

Comparing cosmic growth vs cosmic expansion is one of the premier 
methods for probing the nature of dark energy. Moreover, the details 
of cosmic growth can test the laws of gravity in the universe, on 
scales much greater than solar system or astrophysical tests. Given 
a well defined theory, such tests are straightforward. However, without 
a compelling theory -- not just a class but with a particular functional 
form and hyperparameters -- the comparison with data is more difficult, 
or at best model dependent. 

Allowing the data to play a central role, we demonstrate a model 
independent approach. We find that only two (or in specific physical 
cases three) parameters in the form of binned values of $\gm(a)$ 
deliver subpercent accuracy in fitting the predominant redshift 
space distortion observable $\fs(a)$. This extends to all redshifts the 
previous high redshift parameter method of Paper 1. 

We stress tested the approach against a set of six, diverse modified 
gravity classes with a variety of time dependences, including DGP 
gravity and $f(R)$ gravity. Residuals against exact behaviors of 
the observable successfully achieved subpercent accuracy. Minimizing 
the residual determines the bin values, while any remaining pattern 
offers concrete guidance to the need for a third bin or not. 

We propagated the residuals from the parametrization to cosmological 
parameter bias and showed they are negligible, at below the effective 
$0.1\sigma$ level in joint confidence contours, for next generation 
data of the characteristics of the DESI galaxy redshift survey. 

As importantly, the method lays out a clear path for interpreting the 
bin parametrization results in terms of the physical signature of the 
cosmological gravity theory. Based on the trend of values, their 
steepness, magnitude, and any need for an early bin or scale dependence, 
this approach can guide the search for the laws of cosmic gravity in 
the appropriate direction. 

Future work includes whether such a method can be fruitful for weak 
gravitational lensing: like cosmic growth it relies on a modified Poisson 
equation, with $G_{\rm light}(a)$ instead of $\gm(a)$, but with a 
different kernel. If it too can be parametrized for the observables 
in a low dimensional manner, then next generation surveys will -- even 
in a model independent manner -- have 
excellent capabilities to explore cosmic gravity.

\acknowledgments 

This work benefitted from discussions during the Energetic Cosmos 
Laboratory conference 
``Exploring the Energetic Universe 2017'' at Nazarbayev University. 
EL is also grateful to the Yukawa Institute for Theoretical Physics at Kyoto 
University for hospitality during YITP workshop YITP-T-17-03, and useful 
discussions, especially with Kazuya Koyama. 
This work is supported in part by the Energetic Cosmos Laboratory and by 
the U.S.\ Department of Energy, Office of Science, Office of High Energy 
Physics, under Award DE-SC-0007867 and contract no.\ DE-AC02-05CH11231.


\end{document}